\documentclass[a4paper,fleqn,usenatbib]{mnras}

\usepackage[T1]{fontenc}
\usepackage{ae,aecompl}

\usepackage{graphicx}	% Including figure files
\usepackage{amsmath}	% Advanced maths commands
\usepackage{amssymb}	% Extra maths symbols

\newcommand{\Kepler}{{\it Kepler}}
\newcommand{\ms}{\ensuremath{\rm m\,s^{-1}}}
\newcommand{\bv}{$B - V$}

\title[RV Detection Biases from Stellar Rotation]{Radial Velocity Planet Detection Biases at the Stellar Rotational Period}

\author[Vanderburg et al.]{Andrew Vanderburg$^{1,2}$\thanks{E-mail: avanderburg@cfa.harvard.edu},
 Peter Plavchan$^{3}$,
John Asher Johnson$^{1}$,
\newauthor David R. Ciardi$^{4}$,
Jonathan Swift$^{5}$,
Stephen R. Kane$^{6}$
\\
$^{1}$Harvard-Smithsonian Center for Astrophysics, Cambridge, MA 02138\\
$^{2}$NSF Graduate Research Fellow\\
$^{3}$Missouri State University, Springfield, MO, 65897\\
$^{4}$NASA Exoplanet Science Institute, California Institute of Technology, Pasadena, CA 91125\\
$^{5}$The Thacher School, 5025 Thacher Rd. Ojai, CA 93023\\
$^{6}$San Francisco State University, San Francisco, CA 94132
}

\pubyear{2016}

\begin{document}
\label{firstpage}
\pagerange{\pageref{firstpage}--\pageref{lastpage}}
\maketitle

% Abstract of the paper
\begin{abstract}
Future generations of precise radial velocity (RV) surveys aim to achieve sensitivity sufficient to detect Earth mass planets orbiting in their stars' habitable zones. A major obstacle to this goal is astrophysical radial velocity noise caused by active areas moving across the stellar limb as a star rotates. In this paper, we quantify how stellar activity impacts exoplanet detection with radial velocities as a function of orbital and stellar rotational periods. We perform data-driven simulations of how stellar rotation affects planet detectability and compile and present relations for the typical timescale and amplitude of stellar radial velocity noise as a function of stellar mass. We show that the characteristic timescales of quasi-periodic radial velocity jitter from stellar rotational modulations coincides with the orbital period of habitable zone exoplanets around early M-dwarfs. These coincident periods underscore the importance of monitoring the targets of RV habitable zone planet surveys through simultaneous photometric measurements for determining rotation periods and activity signals, and mitigating activity signals using spectroscopic indicators and/or RV measurements at different wavelengths. 
\end{abstract}


\section{Introduction}

Over the last thirty years, stellar radial velocity (RV) measurements have pushed to increasingly high precision \citep{Campbellwalker, 3ms-1, elodie}. As RV precision improved further in the 1990s and 2000s to 1-2 \ms\ precision (per single measurement) with new instruments like the HIRES spectrograph on the Keck I telescope \citep{hires} and the HARPS spectrograph on the ESO 3.6 meter telescope \citep{harps}, studies of stellar jitter became more common \citep{saarjitter, wrightjitter, isaacsonjitter, dumusquespots,dumusquegranulation, bastienjitter}. Stellar activity can cause radial velocity variations on a variety of different timescales. Asteroseismic oscillations in typical main-sequence solar-mass stars can induce RV variations of order 1 ms$^{-1}$ on timescales of minutes \citep[e.g.][]{butleralphacen}. Granulation on the stellar surface has also been shown to cause significant radial velocity jitter, in the 1-3 ms$^{-1}$ range on the timescale of hours to days \citep{dumusquegranulation}. 

Stellar activity can induce RV variations on longer timescales as well. Starspots can give rise to RV variations by modifying the spectral line profile as the starspot moves across the stellar disk \citep[e.g.][]{quelozspots}. These modulations can be quasi-periodic at the rotation period (and its harmonics) of the star, which can range from hours to months depending on the spectral type and age. The RV modulations from starspots are further complicated by the lifetime of spots and differential stellar rotation, leading to time variable phases and amplitudes of the RV variations. Even longer period RV oscillations can be caused by stellar magnetic activity cycles, which have typical timescales of several years to decades \citep{harpsactivityvelocity}. 

RV jitter from stellar activity can impede and confuse searches for Keplerian radial velocity variations from orbiting planetary companions \citep{andersen}. In some cases, particularly for short timescale variations from granulation and asteroseismic oscillations, RV jitter from stellar activity can be treated as a Gaussian random error added in quadrature to measurement errors \citep[e.g.][]{gregory05}. However, longer period RV variations can be more difficult to treat because their characteristic timescales overlap with the orbital periods of many exoplanets. Techniques such as harmonic modeling have shown potential to mitigate noise from long timescale processes given certain conditions \citep[e.g.][]{howardk78, pepek78, corot7bhatzesmass}. In some cases, it is difficult to distinguish a planetary RV signal from activity, for example when stellar rotation periods are close to the orbital period of a planet candidate \citep[e.g.][]{dragomir}.

It is possible but challenging to detect planets when their Keplerian RV amplitude is significantly smaller than the amplitude of quasi-periodic RV noise caused by stellar activity. Two examples of this are Kepler 78-b \citep{sok78, howardk78, pepek78, hatzesk78, grunblatt}, and Corot 7-b \citep{corot7b, corot7bhatzesmass,haywoodcorot7}. In these cases, the activity timescales of the host stars were at least an order of magnitude longer than the planetary orbital periods, making it possible to effectively high-pass filter out stellar activity to recover the planetary masses. Detecting these two planets with RVs is somewhat simplified because Kepler 78-b and Corot 7-b transit their host stars and have well determined orbital periods and times of conjunction, although it is possible to detect the signals without this prior knowledge \citep{hatzesk78, faria}. Alternative approaches to recovering small amplitude planetary signals include decorrelation against activity-sensitive spectroscopic indicators such as the Ca II H \& K lines, the Ca II infrared triplet \citep{catriplet} and the H-$\alpha$ line \citep{robertson1}, although presently even the best activity indicators are not perfect.

Detecting small amplitude planetary signals and disentangling them from stellar activity becomes more difficult when the orbital period of the planet is close to that of the stellar activity. Recently, this has been illustrated by the realization that several RV signals attributed to low mass exoplanets orbiting ostensibly quiet M-dwarfs \citep[GJ 581, GJ 667, and Kapteyn's star,][]{gj581disc,gj667c, kapteyndisc} are likely the result of low amplitude stellar RV variability \citep{robertson1, robertson2, kapteyn}. These studies used measurements of spectroscopic activity indicators (in particular, the H-$\alpha$ indicator) to both determine the host star's rotation period and search for correlations between velocities and RV signals. In these three systems, the authors found correlations between activity indicators and the RV signal previously attributed to low mass exoplanets. While some of these claims are still disputed \citep[e.g.][]{comment}, it is clear that properly disentangling the RV signal of planets from the signals caused by stellar activity is crucial for detecting small planets with radial velocities. Stellar activity's impact on RV measurements is an active area of research, and new techniques, activity indicators, and instruments are being developed to mitigate this problem.

In this paper, we quantify the impact that stellar activity signals can have on the detection of planets in different orbital periods. In Section \ref{simulations}, we simulate the detection of planets in the presence of realistic stellar radial velocity noise and map out the sensitivity (or lack thereof) to planets in different orbital periods. We compile and combine empirical relations in Section \ref{analytical} to investigate how regions of high and low sensitivity scale with stellar mass, and validate it in Section \ref{verification} using simulations based on photometric data from the \Kepler\ space telescope. In Section \ref{spurious}, we investigate how uncorrected or poorly corrected stellar activity signals can mimic exoplanets, and in Section \ref{discussion}, we compare the stellar activity timescales and habitable zone orbital periods and identify the range of stellar masses most amenable to detecting potentially habitable exoplanets through RV measurements in the presence of stellar activity induced RV jitter. 

\section{Impact of Correlated Activity Noise on Planet Detection} \label{simulations}

Previous studies \citep{boisse11, robertson1,robertson2, howardgl15a} have pointed out that it is difficult to detect RV planets in orbits on the timescale of the stellar rotation period. In this section, we quantify the difficulty in detecting planets as a function of orbital period in the presence of rotationally induced RV jitter. We do this by synthesizing stellar activity RV time series using precise photometric data, injecting planets at a range of orbital periods, and testing how stellar noise affected the detectability of the signal. These tests simulate a best-case scenario for removing stellar activity signals in terms of instrumental RV precision and stability, dense sampling and cadence, and usefulness of spectroscopic indicators. We show that even in ideal circumstances, biases exist which inhibit the detection of exoplanets orbiting near the stellar rotation period and its harmonics.  

We started by simulating the presence of stellar activity in the radial velocity time series using the FF' method developed by \citet{aigrainspots} on \Kepler\ data. We chose several stars with high amplitude rotational variability (necessary to precisely calculate the time derivative) and fitted the \Kepler\ light curve with a basis spline (B-spline) with breakpoints every 1.5 days. We iteratively excluded 3--$\sigma$ outliers and refit the spline until convergence. We then used the spline fit light curve to calculate the derivative of the time series, which we then multiplied by the time series itself to calculate the expected radial velocity. We set the amplitude of the starspot RV signal using the approximate prescription given by \citet{aigrainspots}, which yielded semi-amplitudes typically between 2 and 15 \ms, which are typical of active stars with high amplitude photometric variability like those we have selected. We note that plages and faculae, which may cause significant RV variations, often don't leave a photometric signature in a white light curve, so the FF' method does not take these signals into account \citep{haywood16}. The timescales and periodicities of these RV variations are the same as for starspots, so including them would not significantly change the results of this analysis.

We then simulated an idealized, high cadence observing schedule on which to acquire radial velocity measurements. We simulated nightly observations that occurred randomly throughout the night for 100 nights. We did not take into account scheduling complications (like weather) that might interrupt the high cadence observations. We added to the stellar RV time series taken from the \Kepler\ light curves a white random noise term, with a dispersion of 20\% the standard deviation of the stellar RV signal time series to simulate photon noise and/or instrumental jitter. This ensures we have sufficient precision to resolve the stellar RV signal. The typical measurement uncertainties range from 0.5 to 3 \ms, consistent with or larger than measurement uncertainties expected for current \citep{eprv} and future generations of RV spectrographs \citep[][]{espresso}.

We also simulated measurements of a generic stellar activity indicator, that correlates with the measured RV. This could in principle represent a combination of a measurement of the calcium or H-$\alpha$ indices, line shape diagnostics like the bisector or full width at half maximum, or other indicators \citep[e.g. ][]{figueira}. We assume that there is a simple linear relation between our generic activity indicator and the measured stellar RV, and added a dispersion in the activity indicator due to measurement errors such that the total random error in the RVs comes equally from the photon noise in the RV measurements and the random error in the activity indicator. We note that the assumption that there is a linear relationship between the activity indicator and the stellar RV signal is a best-case scenario, and typically the relationship is more complex.

We then injected planetary RV signals into the stellar activity plus photon noise time series we generated previously. For each star, we injected a total of 10000 signals, with periods spaced logarithmically between 1 and 100 days. We injected planets in circular orbits with RV semi-amplitudes half the typical amplitude of the stellar activity signal and with randomly distributed orbital phases. 

After generating the simulated RV time series, we began analysis to attempt to mitigate the stellar RV signal and recover the planets. We fit a line to the measured stellar RV signal with the planet injected and activity indicator, and subtracted off the correlation. This approach of sequentially subtracting off the best-fit stellar activity signal could potentially dilute a planet's signal, but this approach is reasonable for the initial detection of planet candidates, which is what we simulate here. A more sophisticated analysis (see Section \ref{sophisticated}) is typically done after the initial signal detection. Moreover, sequentially fitting for stellar variability and searching for planets in the residuals of this fit is viable in this regime where the peak-to-peak variations in the stellar RV signal are 4-5 times greater than the planet's peak-to-peak RV amplitude. We tested that we were not significantly corrupting the planetary signals by injecting signals of different amplitudes compared to the stellar activity, and found qualitatively similar results for these injections. 

After removing the stellar activity from the time series, we calculated a Lomb-Scargle periodogram \citep{scargle}. We recorded $P_{\rm stellar}$, which we define as the power of the highest periodogram peak within 1\% of the injected planet period. We then calculated a periodogram of the RV time series with only the planetary signal and random noise (and without the stellar activity signal) and repeated the measurement to obtain $P_{\rm white}$, the maximum power within 1\% of the planet period in this periodogram without the stellar noise. We define $R_{\rm S/N}$ to be the square root of the ratio between the Lomb-Scargle power measured with and without the stellar signal included as a measure of detection efficiency.  

\begin{equation}
R_{\rm S/N} = \sqrt{\frac{P_{\rm stellar}}{P_{\rm white}}}
\end{equation}

Because the square root of a Lomb-Scargle power measurement is a linear amplitude, $R_{\rm S/N}$ is equivalent to the ratio of signal-to-noise of a detection with and without the presence of stellar activity . 

\begin{figure*}
%\epsscale{1}
\includegraphics[width=\linewidth]{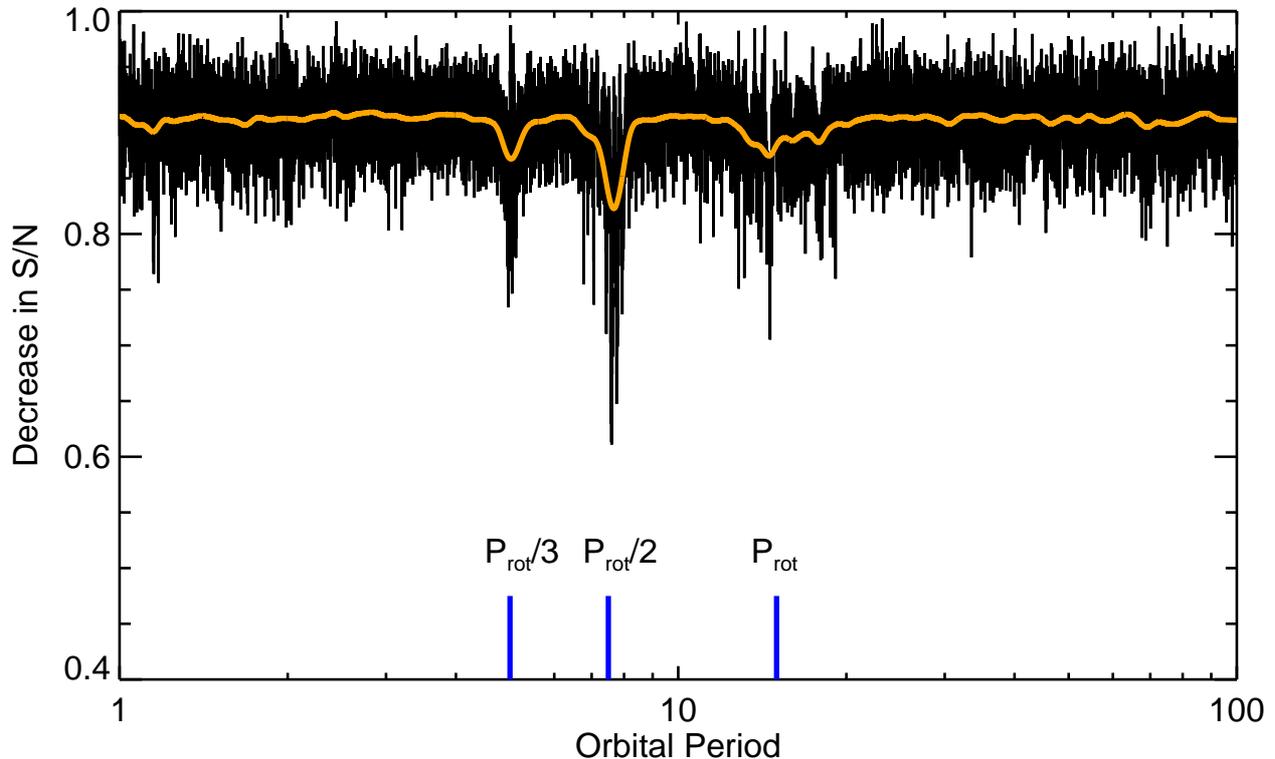}
\caption{Relative detection efficiency ($R_{\rm S/N}$) of planets around KOI 2007 as a function of orbital period. The thick orange line is a Gaussian smoothed version of the individual detection efficiencies for each of the 10000 trials, shown in black. We include blue hash marks at the stellar rotation period and its first two harmonics. The baseline level of detection efficiency is about 0.9, implying that adding stellar activity generally decreases the efficiency of planet detections by about 10\% around this star, under our assumptions. There are three troughs at orbital periods corresponding to the rotation period of star and its first two harmonics. Even under idealized circumstances, it is more difficult to find planets with orbital periods close to the stellar rotation period and its harmonics. }\label{relativepower}
\end{figure*}

In Figure \ref{relativepower}, we show the $R_{\rm S/N}$ for planets injected into one particular star, an active planet host called KOI 2007. For this particular star, in general, the stellar activity and correction decreases the efficiency of planet detections to about 90\% the efficiency of planet detection in the case of no stellar activity. Near the star's rotation period and its first two harmonics, the detection efficiency decreases from its baseline level. This decrease in detection efficiency is the signature of a degeneracy between between the planet's orbital signal and the stellar activity signal. 

We note that for KOI 2007, the first harmonic of the rotation period shows the greatest decrease in detection efficiency. This effect, which we observe in many but not all stars, is because as a spot moves across the limb of the star, the apparent RV shift undergoes an oscillation on the timescale of half a rotation period. This introduces significant power at one half the rotation period. 

\section{RV Jitter from Rotational Modulation}\label{analytical}

In this section we investigate how this decreased detection efficiency at the rotation periods and its two harmonics scales with stellar mass. We calculate relationships between the typical timescale and amplitude of RV variations due to stellar rotation as a function of stellar mass.

\subsection{Timescales}\label{timescales}

We estimate the typical rotation periods of stars (and therefore, the typical period of starspot induced RV variation) using an empirical gyrochronology relation of \citet{barnes}, which relates the age, rotation period, and \bv\ with a typical dispersion of 15$\%$. We relate the age, stellar mass, and \bv\ using stellar evolution models from the Dartmouth Stellar Evolution Database \citep{Dotter08}. These models provide relations between a host star's mass, radius, luminosity, effective temperature, and colors. We use isochrones for stars with solar [Fe/H], [$\alpha$/Fe], and helium abundance. We obtain the range of possible rotation periods by taking the age to be anything between 1 and 10 Gyr, and assuming a dispersion of 30$\%$ (or roughly two $\sigma$). 

We also compare the \citet{barnes} gyrochronology relation with the sample of rotation periods from \citet{mcquillanrotationall} from the \Kepler\ dataset. This sample differs in detail with the gyrochronology relations, but the general trends are in agreement.

\begin{figure*}
%\epsscale{1}
  \begin{center}
      \leavevmode
\includegraphics[width=\linewidth]{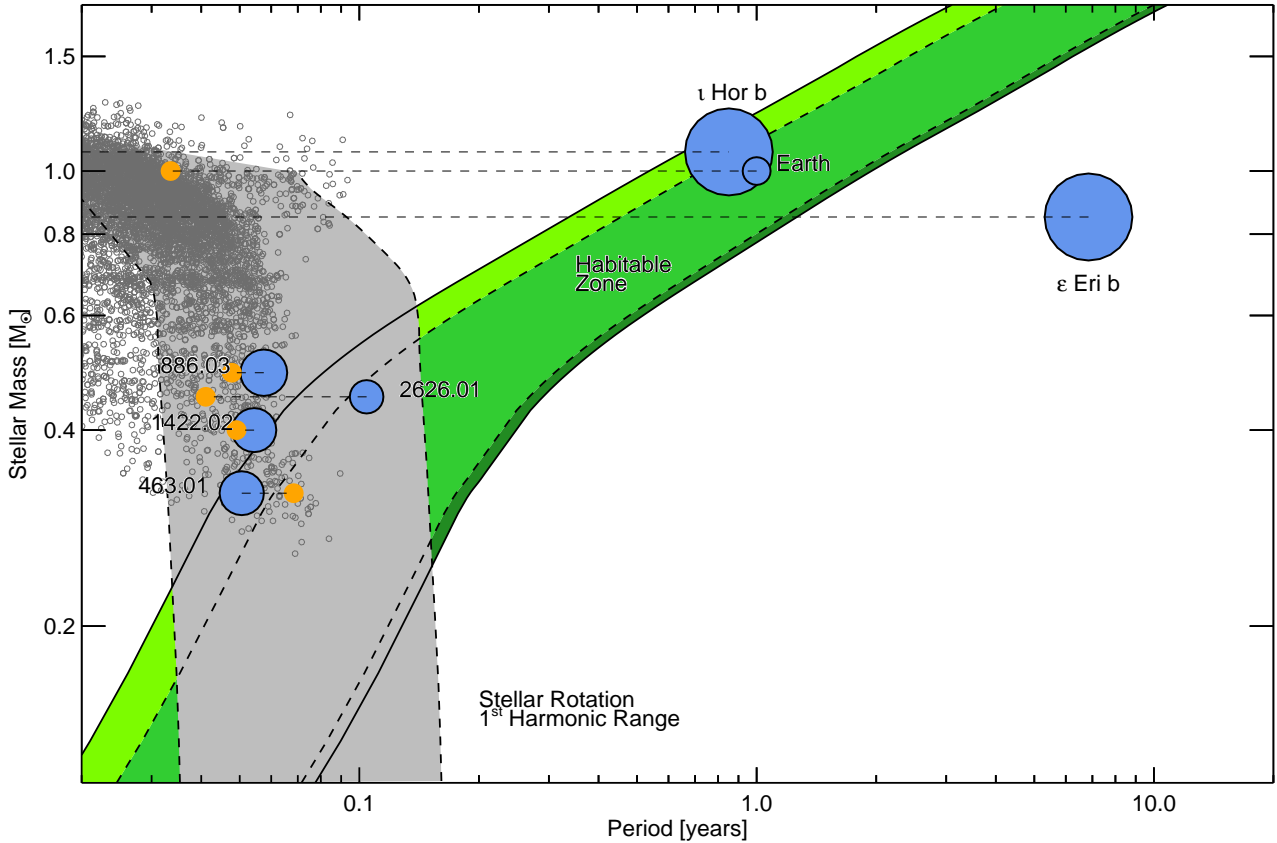}
\caption{Characteristic timescales of periodic or quasi-periodic radial velocity signals due to both stellar activity and planetary companions. The \citet{kopparapu} inner optimistic, conservative, and outer optimistic habitable zones are shown in light, medium, and dark green, respectively. The range of the first harmonic of rotation periods for main-sequence stars between 1 and 10 Gyr are shown in gray. Selected first harmonics of rotation periods for planet host stars are shown as orange dots. The numbers labeling some of these stars are \Kepler\ objects of interest designations.The planets' orbital periods are shown as blue dots (with the size of the dot roughly corresponding to the planetary radius), and are connected to their hosts' half stellar rotation periods with thin dashed horizontal lines. A random subset of the first harmonic of rotation periods measured by \citet{mcquillanrotationall} are also plotted in dark grey octagons. This empirical sample of rotation periods is somewhat discrepant from the gyrochronology relations, but shows similar trends. $\epsilon$ Eridani and $\iota$ Horologii are young stars with ages of less than 1 Gyr \citep{epsiloneri, iotahorpcyc}, explaining their fast rotation.} \label{massvsperiod}
\end{center}
\end{figure*}

We found in Section \ref{simulations} that RV noise is introduced predominantly at the stellar rotation period and the first two harmonics. This was previously noted by \citet{aigrainspots} and \citet{boisse11}, who showed that the RV noise is typically introduced at the rotation period and its first two harmonics, and that the first harmonic is dominant. We similarly find that for most stars, the highest peak in a periodogram of the FF'-estimated RV signal is at the first harmonic (see Section \ref{spurious}), and when we repeat our analysis described in Section \ref{simulations} for a large sample of stars, the reduction in detection efficiency is most often greatest near the first harmonic. We therefore take the first harmonic (or one half of the orbital period) to be the timescale of the greatest disruption to RV detection efficiency. We plot the orbital periods affected by stellar rotation versus stellar mass in Figure \ref{massvsperiod}. The gyrochronology relations show that the orbital periods affected by noise from stellar rotation decrease with increasing stellar mass. For comparison, we overplot the orbital period of exoplanets in their host stars' habitable zones, and we include (and label) some example planets and their host stars' rotational periods.

\subsection{Amplitudes}

We calculate the typical amplitude of rotational radial velocity variations due to starspots. The physical mechanism behind these variations is well understood -- as cool spots move across the limb of the star, they introduce an asymmetry in the rotational velocity profile of the star's disk and therefore introduce an asymmetry in the spectral line shape. Although RV variations can also be caused by plages and faculae, the effects of these active groups are more difficult to quantify given only a light curve \citep{haywoodcorot7} and are only the dominant cause of RV variations for relatively inactive slowly rotating stars \citep{haywood16}. Therefore we focus on starspots in this analysis. \citet{aigrainspots} give an analytic model for a spot induced RV time series given a flux time series, but for our analysis, we simplify their relation to estimate peak-to-peak amplitudes: 

\begin{equation} \label{spotrv}
RV_{pp} \simeq F_{pp} \times v \sin(i),
\end{equation}

\noindent where $RV_{pp}$ is the peak-to-peak RV variation caused by starspots, $F_{pp}$ is the peak-to-peak flux variation in the passband of the RV measurements, and $v\sin(i)$ is the projected rotational velocity of the star. This estimate holds when the flux variations are measured over roughly the same bandpass as the radial velocities. Our estimates will focus on optical flux variations (as measured by the {\em Kepler} telescope; NIR flux variations (and hence RV variations) are likely lower due to lessened flux contrasts with cool starspots \citep{reinersspots}. 

We estimate the typical magnitude of the peak-to-peak variations in optical flux caused by starspots from the results of \citet{mcquillanrotationall}, who measured the periodic photometric amplitude variations ($F_{pp}$) of {\em Kepler} targets. They defined the periodic amplitude as the range between the 5$^{th}$ and 95$^{th}$ percentile of median divided, unity subtracted, and 10-hour boxcar-smoothed {\em Kepler} light curves.  This treatment suppresses variation on timescales longer than a {\em Kepler} observing quarter and shorter than ten hours. After this treatment, the dominant source of photometric variability is starspot modulation. 

We use the results of \citet{mcquillanrotationall} to estimate a relationship between rotational modulation and stellar mass. We divide the data into bins of size 0.1 $M_{\sun}$ (using mass estimates from  \citealp{mcquillanrotationall}), and take the 16$^{th}$ and 84$^{th}$ percentile (roughly one $\sigma$) of each mass bin as lower and upper bounds for typical rotational modulation amplitudes. This sample of {\em Kepler} target stars includes very few stars with mass less than 0.3 $M_{\sun}$, so we extrapolate our relationships to lower mass M-dwarfs.  The \citet{mcquillanrotationall} sample excludes stars for which no rotation period was securely detected, and is therefore incomplete for stars with longer rotation periods and lower rotational amplitudes. To see if this bias affects our estimate significantly, we verified our relationship using measurements of photometric variability from \citet{basrirange}, which includes stars for which no rotation period was detected. The measurements from \citet{mcquillanrotationall} and \citet{basrirange} show the same trend towards larger photometric variations in low-mass stars (M $\lesssim 0.5 M_{\sun}$) and report variations of roughly the same amplitude.

We estimate typical values of $v\sin(i)$ by combining our estimates of rotation periods for main-sequence stars from Section \ref{timescales}, stellar radii from a 5 Gyr Dartmouth isochrone, and the average value of $\sin(i)$ (over all possible spin axis orientations) using: 

\begin{equation}\label{vsini}
<v \sin(i)> \simeq \frac{2 \pi R_{\star}}{P_{rot}} \times <\sin(i)>,
\end{equation}

\noindent and

\begin{equation}
<\sin(i)> \simeq 0.79,
\end{equation}

\noindent where $R_{\star}$ is the stellar radius and $P_{rot}$ is the stellar rotation period.

\begin{figure*}
%\epsscale{1}
\includegraphics[width=\linewidth]{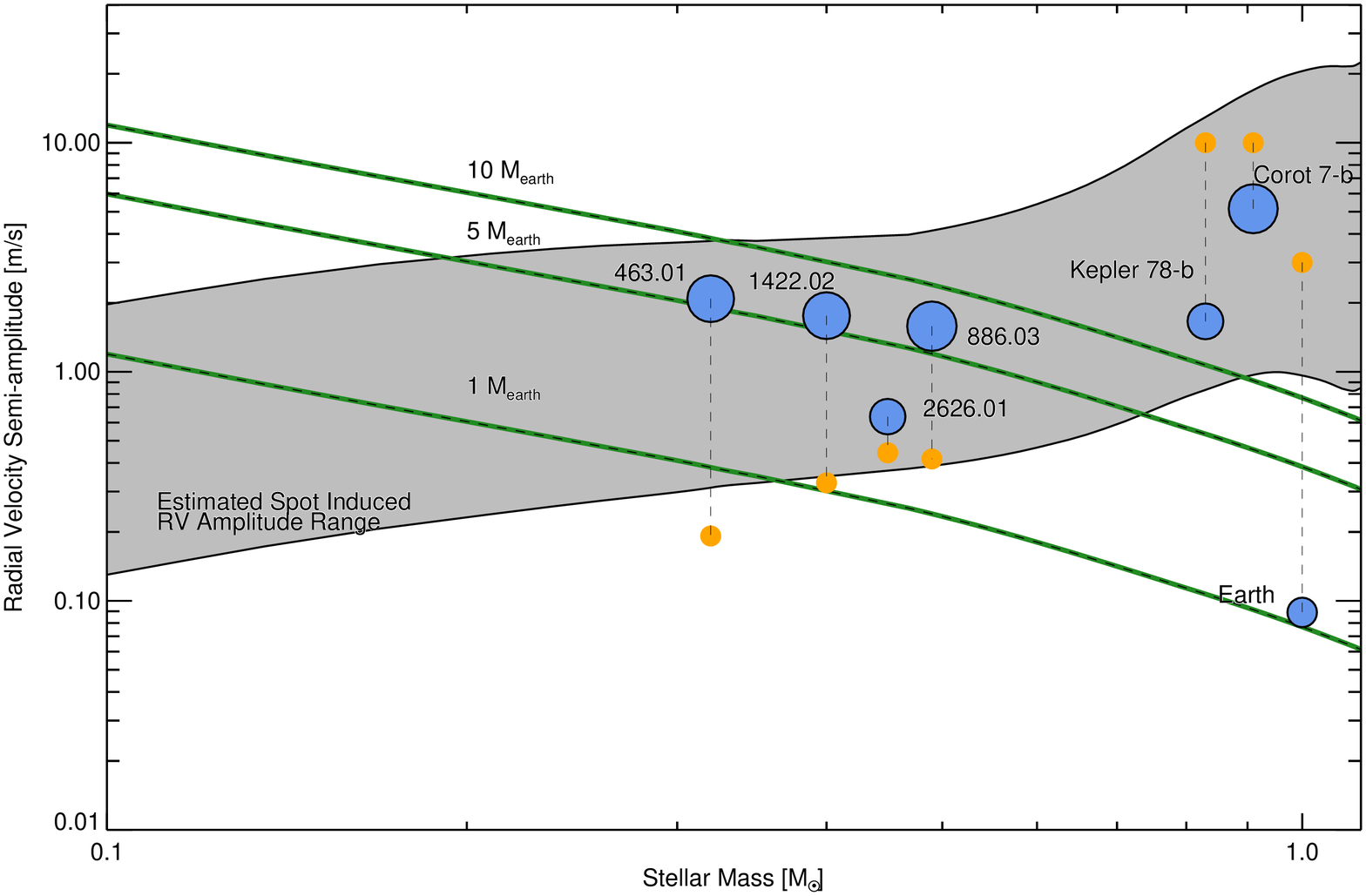}
\caption{Estimated characteristic semi-amplitudes of oscillatory radial velocity signals. The semi-amplitude of rotational modulations due to starspots as a function of stellar mass are shown in grey. The rotational modulations are calculated assuming uniformly distributed stellar inclinations. We also plot the RV semi-amplitude of exoplanets in the habitable zone of stars of various masses in green lines. Characteristic exoplanets are plotted as blue dots, with the size of the dot corresponding to the planetary radius. The numbers labeling some of these planet host stars are \Kepler\ objects of interest designations. Each exoplanet plotted is connected by a thin dashed vertical line to an orange dot, corresponding to the predicted or observed RV semi-amplitude induced by stellar rotation. The typical amplitude of RV variations decreases for lower mass stars because the typical projected rotational velocity decreases substantially with stellar mass. The four Kepler candidates shown have lower photometric amplitudes than typical M-dwarfs, hinting at observational biases in Kepler data against detecting small planets in long period orbits around photometrically noisy stars. The point for Earth falls slightly above the line for an Earth-mass planet in the habitable zone because Earth orbits at the very inner edge of the \citet{kopparapu} habitable zone.}\label{rvvsmass}
\end{figure*}

Combining these relations with Equation \ref{spotrv} gives an estimate of rotational radial velocity modulations as a function of stellar mass. We show our relations in Figure \ref{rvvsmass}. We find that over all stellar masses, the typical amplitude of starspot induced RV signals is of order 1 \ms or larger, which will be relevant for future generations of RV planet searches. We note that because stellar magnetic activity and the amplitude of photometric variations are inversely correlated with the stellar rotation period \citep{mcquillanrotationall}, the amplitude of RV variations will depend even more strongly on the rotation period than Equations \ref{spotrv} and \ref{vsini} indicate.
 
\section{Empirical Verification of Scaling Relations}\label{verification}

We apply the technique described in Section \ref{simulations} to a larger set of stars observed by Kepler to recover an empirical map of $R_{\rm S/N}$ at different orbital periods as a function of stellar mass. We started by choosing stars observed by \Kepler\ from \citet{mcquillankoi} that showed high levels of stellar variability, well measured rotation periods, and were identified as dwarf stars in the \Kepler\ input catalog. We choose stars with high amplitude variability because the FF' method involves taking a time derivative which is difficult to do with low signal-to-noise data. In total, we selected 648 stars, with spectral types ranging from F to M.

\begin{figure*}
%\epsscale{1}

\begin{minipage}{.5\textwidth}
  \centering
  \includegraphics[width=\linewidth]{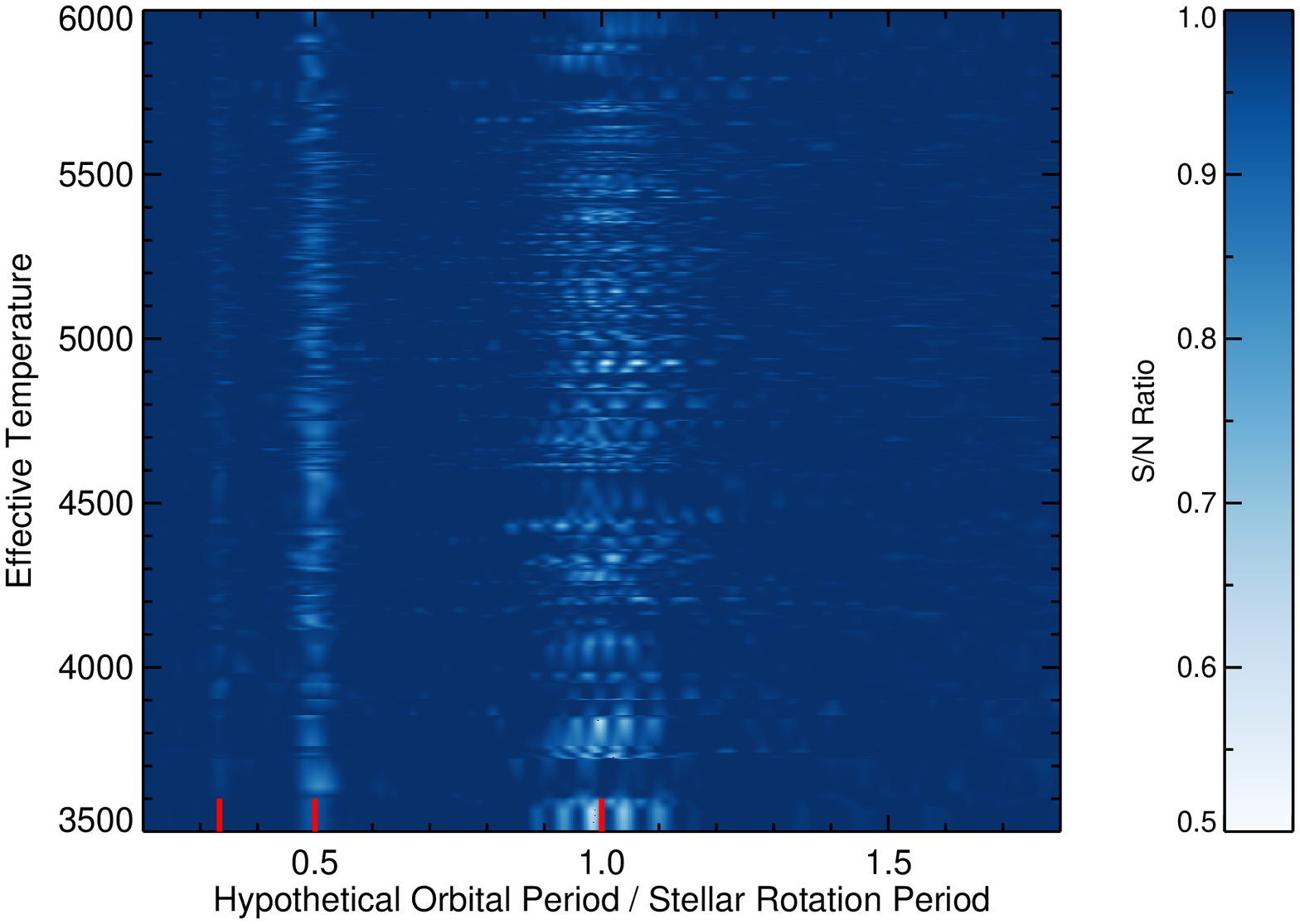}
\end{minipage}%
\begin{minipage}{.5\textwidth}
  \centering
  \includegraphics[width=\linewidth]{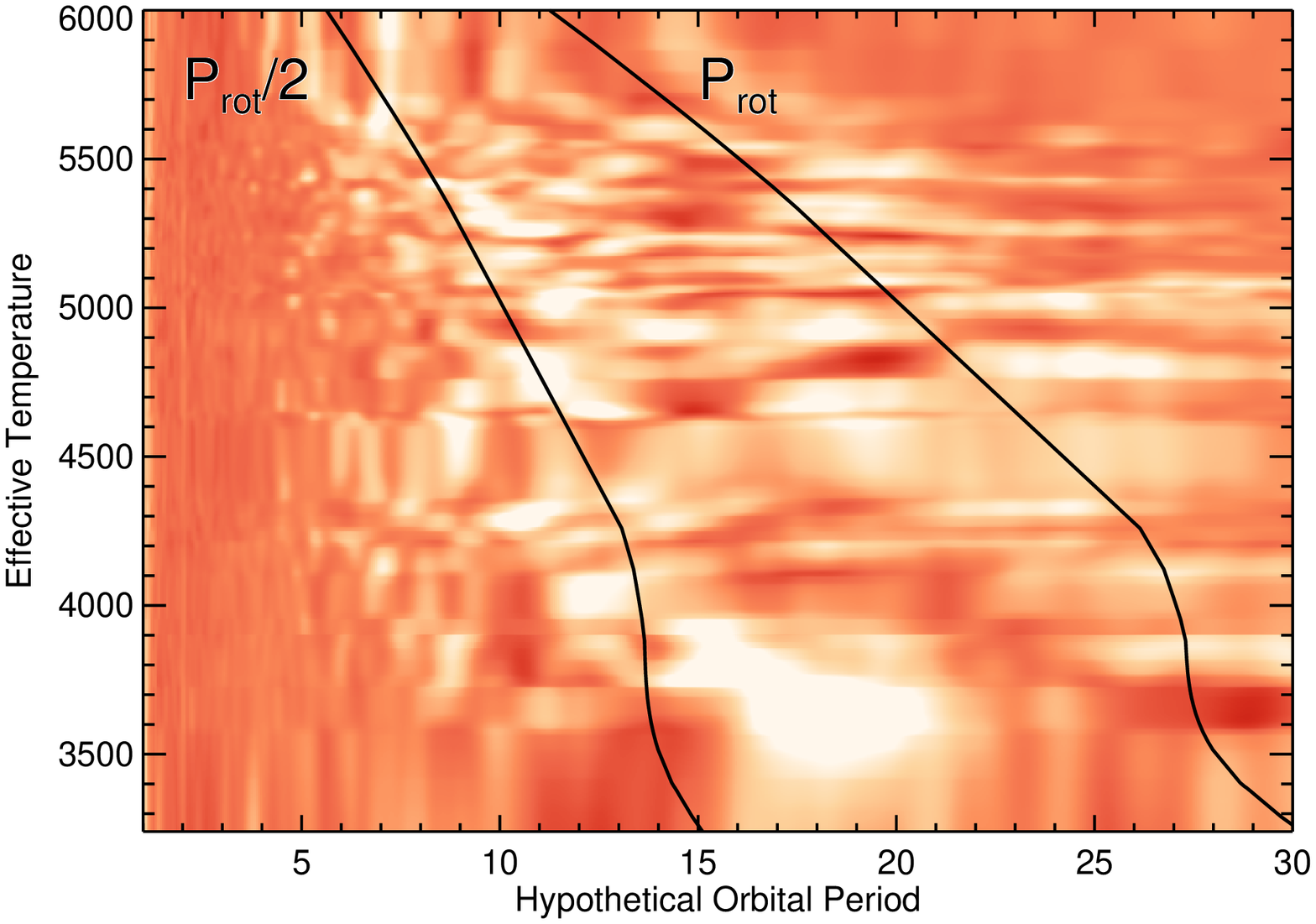}
\end{minipage}

\caption{Reduction in detection efficiency ($R_{\rm S/N}$) of planets with host stars of varying effective temperatures. Left: $R_{\rm S/N}$ for planets at periods relative to the stellar rotation period. Even with optimistic assumptions about the ability to correct for stellar activity induced radial velocity variations, the ability to detect exoplanets is diminished by up to 50\% near the stellar rotation period and its harmonics (shown with red hash marks at the bottom of the panel). Right: $R_{\rm S/N}$ for planets at various orbital periods. There is a band of reduced detection efficiency corresponding to typical stellar rotation periods (and their harmonics) at each effective temperature. The two black lines show the predicted stellar rotation periods and their first harmonics for stars with an age of 1 Gyr.}\label{detectabilityteff}
\end{figure*}

We then applied the same procedure we performed in Section \ref{simulations} to each star's activity time series. We synthesized an RV time series, injected planets, corrected for stellar activity, and measured $R_{\rm S/N}$. Then, we found the typical detection efficiency after injecting the stellar noise into the light curves over all orbital periods, and normalized that efficiency to 1. The presence of stellar activity decreases the detection efficiency for planets at all periods, but we wish to compare the detection efficiency at orbital periods far from the stellar rotation period to the efficiency at orbital periods near the rotation period and its harmonics. Finally, we sorted the stars by their effective temperature, as reported in the Kepler Input Catalog \citep[KIC][]{kic}, and plotted $R_{\rm S/N}$ versus stellar effective temperature and orbital period as a color map in Figure \ref{detectabilityteff}.

In Figure \ref{detectabilityteff}, we show the results in two different ways. First, we plot the reduction in detection efficiency ($R_{\rm S/N}$) as a function of the effective temperature and of the ratio between the hypothetical orbital period and stellar rotation period. There are strong bands of low detectability at the stellar rotation period and at the first harmonic (one half the rotation period), and a weaker band visible at the second harmonic (one third the rotation period). We also plot the reduction in efficiency as a function of stellar effective temperature and absolute orbital period, and find that there are bands of reduced planet detection efficiency that trace orbital periods corresponding at the stellar rotation period and its first harmonic. To show this, we overplot our the rotation period and first harmonic of the predicted stellar rotation periods for stars of the measured effective temperatures. We find that the band of lower detection efficiency is at relatively short periods, corresponding to the first harmonic of stars with an age of about 1 Gyr. We believe this relatively young age is due to our selection of stars with high amplitude starspot variations (for signal-to-noise considerations). Young stars have both higher amplitude and more rapid starspot modulations.

\section{Spurious Planet Detections from Activity}\label{spurious}

Thus far, we have simulated radial velocity observations taken under conditions ideal for removing stellar activity. In particular, we have assumed that we are able to perfectly model and correct for stellar radial velocity variations. Unfortunately, methods for performing this type of stellar activity correction are still maturing and at the present are imperfect. In this section, we demonstrate that inadequate modeling of systematics can lead to spurious planet detections.

\begin{figure}
%\epsscale{1}
  \begin{center}
      \leavevmode
\includegraphics[width=\linewidth]{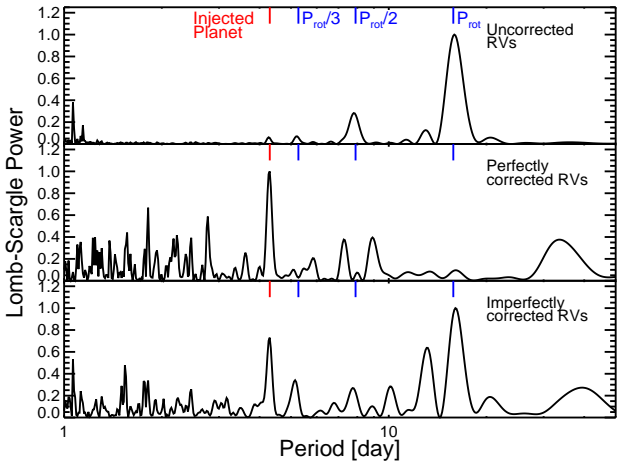}
\caption{Lomb-Scargle periodograms of simulated RV time series from an active star (KOI 254) with an injected planetary signal. Top: periodogram of the RV time series with no correction applied for stellar activity. Activity at the stellar rotation period and its harmonics (marked with blue hash marks at the top of the panel) dominate the signal from the planet (marked with a red hash mark at the top of the panel). Middle: periodogram of the RV time series with an activity correction which models the stellar variations perfectly. The planetary signal is visible in this case, and dominates the remnants of the stellar activity signal. Bottom: periodogram of the RV time series with an activity correction where stellar RV variations are not modeled perfectly. This correction leaves in significant power near the rotation period and its harmonics, which is more significant than the planetary signal. This can lead to spurious planet detections.} \label{imperfect}
\end{center}
\end{figure}

Figure \ref{imperfect} shows Lomb-Scargle periodogams of simulated RV measurements from an active star (KOI 254) with a small planetary RV signal injected. Without any mitigation of the stellar RV signal, the planet is undetectable and dwarfed by the power at the stellar rotation period. When the ``perfect'' correction we have considered so far in this work is applied, the planetary signal is strongly detected, and the stellar signal is largely (but not perfectly) removed due to the random noise we added to the RV curve and indicators. We also simulated an ``imperfect'' correction case. Instead of assuming a linear relationship between the activity indicator and the stellar radial velocity, we assumed a weak power law relationship (with the indicator $\propto$ RV$^{1/4}$). We then attempted to remove the stellar signal by assuming a linear relationship. This type of insufficient modeling is also analogous to scenarios where quasi-periodic stellar activity is modeled as strictly periodic functions \citep[e.g.][]{alphacenbb, rajpaul}. The imperfect removal of the stellar signal greatly decreases the significance of the peaks at the stellar rotation period and its harmonics, but they still contribute significant power and dominate the planetary signal. Spurious RV detections due to stellar activity can still be a problem even when measures are taken to correct for activity induced RV variations.

\begin{figure}
%\epsscale{1}
  \begin{center}
      \leavevmode
\includegraphics[width=\linewidth]{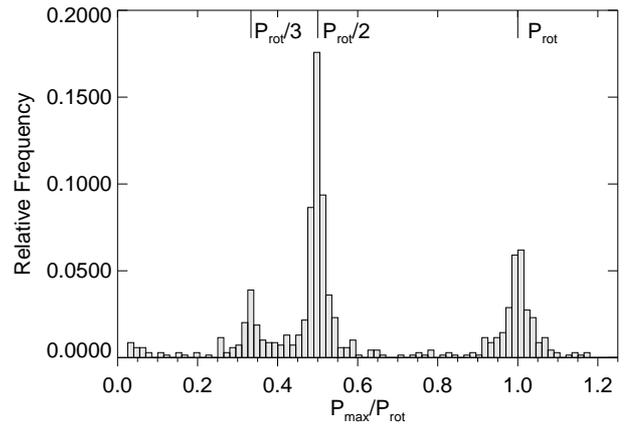}
\caption{Histogram of the ratio between the period at which RV activity is present and the stellar rotation period. The stellar rotation period and its first two harmonics are shown with black hash marks at the top of the plot. These are the periods at which spurious RV detections of planets due to stellar activity will be common. We find that most spurious RV detections will come at harmonics of the stellar rotation period and that the spurious RV detections can be up to 10\% away from the actual harmonic.} \label{activityhistogram}
\end{center}
\end{figure}

We investigated the periods at which these spurious signals tend to appear. We simulated the expected stellar RV signal for the set of active stars as discussed previously in Section \ref{verification} and calculated the Lomb-Scargle periodograms over a range of frequencies from 1 day to 50 days. We then recorded the period of the most significant peak in each periodogram and compared it to the star's rotation period. A histogram of these results is shown in Figure \ref{activityhistogram}. We find that spurious RV signals from stellar activity most often fall at the first harmonic of the rotation period, but the activity on a significant number of stars leads to spurious signals at the rotation period and its second harmonic as well. Interestingly, the peaks are broad about the rotation period, indicating that even RV signals up to 10\% away from the measured rotation period could be caused by stellar activity. These two conclusions are consistent with previous claims of spurious RV detections. \citet{robertson2} claimed that GJ 667C's 105 day rotation period gave rise to a 92 day signal in the RVs, slightly more than 10\% away from the rotation period. Moreover, claims from \citet{robertson1} and \citet{kapteyn} that signals near harmonics of the stellar rotation periods of GJ 581 and Kapteyn's star are due to activity are consistent with our finding that a large number of spurious signals appear at rotation period harmonics.

Finally, we note that with less ideal sampling than we simulate here, the periods at which spurious signals appear can be more difficult to predict due to aliases between the activity signals and the sampling window function \citep[e.g.][]{rajpaul}.

\section{Discussion}\label{discussion}

\subsection{Habitable Zone Exoplanets}

In this paper, we have presented simulations and approximate scaling relations which demonstrate biases against detecting planets at or near the planet host star's rotation period. The simulations show that even under idealized, best-case conditions, it is more difficult to detect exoplanets orbiting near the rotation period of the star. Furthermore, in less ideal circumstances, insufficiently corrected or removed stellar activity can lead to spurious planet detections.

These challenges could hinder efforts to detect and measure the mass of exoplanets in the habitable zones of their host stars, a major goal of future ground based high precision radial velocity surveys. The low (sub-ms$^{-1}$) RV semi-amplitudes of these planets can be dwarfed by RV variations from stellar activity. The challenge of detecting small habitable-zone planets is particularly vexing when the orbital period of a habitable zone exoplanet is close to that of the stellar activity signals. The location of the habitable zone depends on stellar mass, as do the timescales and amplitudes of stellar activity.  Since it is easier to identify exoplanets in RV data when the orbital period is significantly different from the period of stellar activity, it is advantageous to look for habitable zone exoplanets around stars with activity cycle periods significantly different from the orbital period of habitable zone exoplanets.

M-dwarfs are seen as promising targets around which to search for habitable zone exoplanets due to the small size of the stars, their brightness in the near infrared, and the shorter habitable zone periods. In particular, the short HZ orbital periods and low stellar masses significantly reduce the instrumental stability, observational baseline, and cadence requirements necessary to detect HZ exoplanets with radial velocity measurements. For this reason, many radial velocity detections of super-Earth sized planets orbiting in their host stars' habitable zones have come around M-dwarfs \citep[e.g. ][]{wolf}. Confusion of HZ planetary signals with activity induced RV variations, however, can be a concern, because for early-type M-dwarfs (the brightest ones in visible-light and most accessible to existing instruments) stellar rotation periods and their harmonics are often similar to the orbital periods of habitable zone exoplanets. These overlapping period ranges have led to several disputed habitable-zone planet detections \citep{robertson1, kapteyn}. 

On the other hand, G-type dwarf stars have stellar activity periods favorably suited for habitable zone exoplanet searches, although the longer orbital periods and smaller RV semi-amplitudes make such detections more difficult. The Sun, for example, has a $\sim$24 day rotation period compared to a $\sim$400 day habitable zone orbital period. The Keplerian RV signals from potentially habitable planets around sun-like stars are separated from activity signals by an order of magnitude in period, which eases the filtering necessary to detect and extract the Keplerian signals. Therefore, using an optimal observing cadence to detect and avoid the effects of stellar modulation \citep[e.g.][]{dumusquespots, alphacenbb} may be sufficient to find habitable zone exoplanets. Habitable zone exoplanet searches can be fruitful at optical wavelengths (the peak wavelengths of G-stars), even as several new instruments are being built to search for planets in the near infrared around M-dwarfs. A possible drawback for searching for planets around G-dwarfs is that plages and faculae are more important for these stars, adding a new source of activity to be corrected before detecting small planets.

\subsection{Sophisticated Modeling Techniques}\label{sophisticated}

One proposed way to overcome the problem of stellar activity contaminating RV measurements is to perform more sophisticated and complex statistical analyses than the ones considered in this work. In particular, techniques like Bayesian model comparison \citep{nelson} combined with simultaneous modeling of planetary and stellar signals \citep{angladaescude} could help resolve confusion between stellar and planetary signals. However, there are challenges with these approaches. First, there is no one easily invertible RV correlated activity indicator like the one we considered in this work. It could be possible to solve these challenges using physical models \citep{soap2, dumusqueinclination, herrero} or approximate relations \citep{aigrainspots, rajpaul} to work out the expected stellar RV signal, but high precision recovery of the additional planetary RV signal is difficult (Dumusque et al. 2016 in prep). Sometimes, one particular indicator \citep[like H-$\alpha$, ][]{robertson1} correlates with radial velocity, but in general, the relationships are more complex. 

While simultaneous fitting in principle allows recovery of both the planetary and stellar signals, when these signals are on similar timescales, there will be degeneracies which will decrease the detection significance of any planetary signal (as we see in our simulation results). These degeneracies will make it more difficult to detect low amplitude signals at the cutting edge of instrumental precision. 

\subsection{Multi-wavelength RV measurements}

A combination of optical and near infrared RV measurements will be a powerful tool to vet candidate HZ exoplanets. Unlike Doppler shifts, RV signals from starspot modulation are wavelength dependent, and typically are less noticeable at redder wavelengths \citep{reinersspots, escudesurfing}. Multi-wavelength RVs have been used to vet and refute candidate planetary companions to stars, such as TW Hydrae \citep {setiawan, notwhydrae}. 

Similarly, complementary or simultaneous RV measurements in the optical and near infrared \citep{carmenes, gagne} could also be effective in identifying RV variations from starspot modulations by extending the effective bandpass from optical wavelengths to K-band. The larger bandpass could also make it possible to test the wavelength independence of RV signals, and even model and subtract RV signals from starspots by taking advantage of their wavelength dependence. 

Multi-wavelength RV confirmation will be more difficult for M-dwarfs than for solar mass stars because the Wien tail of the M-dwarf blackbody causes optical flux to rapidly decrease with decreasing stellar mass. Many bright M-dwarfs in NIR will be expensive to follow up with optical measurements due to their faintness.

%Simultaneous NIR and optical measurements from instruments such as CARMENES \citep{carmenes} will even more powerfully suppress wavelength dependent RV variations within a single shot. 

%Although the mechanism by which magnetic activity cycles cause RV variations is less well understood, multi-wavelength RV followup will still be important. One mechanism by which magnetic activity cycles can induce long period RV variations is through movement of starspot groups latitudinally on the star. Because such variations are caused by active areas modifying stellar line profiles, they are wavelength dependent and would be suppressed by multi-wavelength RVs. On the other hand, changes in the photospheric radius of the star could lead to time variable gravitational redshifts, which are wavelength independent \citep{loebredshift}. 

\subsection{Photometric Followup}

Photometric followup of candidate habitable zone exoplanet host stars will be crucial in determining whether a candidate Keplerian RV signal is actually caused by an exoplanet or stellar activity. Ground-based \citep{henryrotation} and space-based photometric followup can identify rotation periods of stars and search for transits, and simultaneous photometric measurements from all-sky high precision photometric surveys like TESS\footnote{\url{http://space.mit.edu/TESS}} and PLATO\footnote{\url{http://sci.esa.int/plato/}}, will be useful to predict RV signals from active areas \citep{aigrainspots}. 

By 2030, stars in parts of the {\em Kepler} field will have been photometrically monitored by three different high precision space-based photometric surveys over the course of 21 years (specifically, by {\em Kepler} from 2009-2013, {\em TESS} for one month between 2017 to 2019, and {\em PLATO} for two years between 2024-2030). The long time baseline of high quality photometry could enable the detection of magnetic activity cycles by measuring changes in the starspot coverage of stars, the same way the Sun's magnetic cycle was originally discovered. 

Future work can focus on combining the various diagnostic tools at (or soon to be at) our disposal, including high quality photometry, multi-wavelength RVs, and activity indicators, to overcome the challenges posed by disentangling stellar activity from Keplerian signals. This type of analysis will become more important as technology develops and RV precision continues to improve in both the optical and near infrared.

\section*{Acknowledgements}
We are grateful to Howard Isaacson and Amy McQuillan for helpful advice. We thank Juliette Becker, Thayne Currie, Jonathan Gagn\'e, Peter Gao, and Angelle Tanner for their helpful comments on an early draft of the manuscript. We thank the referee, Suzanne Aigrain, for a thoughtful and detailed report which significantly improved this work. This research has made use of NASA's Astrophysics Data System, the SIMBAD database, operated at CDS, Strasbourg, France, as well as the NASA Exoplanet Archive, which is operated by the California Institute of Technology, under contract with the National Aeronautics and Space Administration under the Exoplanet Exploration Program. A.V. is supported by the NSF Graduate Research Fellowship, Grant No. DGE 1144152. J.A.J is supported by generous grants from the David and Lucille Packard Foundation and the Alfred P. Sloan Foundation.

%\bibliographystyle{mnras}
%\bibliography{refs} % if your bibtex file is called example.bib

\bsp	% typesetting comment
\label{lastpage}
\end{document}